# Lithium Niobate Resonators for Power Conversion : Spurious Mode Suppression Via an Active Ring

†Vakhtang Chulukhadze*, ‡Eric Allen Stolt, ‡Clarissa Daniel, ‡Juan Rivas-Davila, and †Ruochen Lu
†University of Texas at Austin, Department of Electrical and Computer Engineering, Austin, TX, USA
‡Stanford University, Department of Electrical Engineering, Stanford, CA, USA
*vatoc@utexas.edu, stolt@stanford.edu, cydaniel@stanford.edu, jmrivas@stanford.edu, ruochen.lu@austin.utexas.edu

*Abstract*— In an effort to shift the paradigm of power conversion, acoustic resonators pose as compact alternatives for lossy magnetic inductors. Currently, the acoustic resonator's restricted inductive region between its series and parallel resonances constitutes a major bottleneck, which is further diminished due to spurious modes. Prior work has partially addressed this issue by the introduction of various design guidelines tailored to the material and the mode of interest, but can only provide a limited spurious-free region. Alternatively, a separated grounded ring on LN operating in the first order symmetric lamb mode (S1), maintains optimal device performance with a large fractional spurious mode suppressed region, but has been shown to experience voltage breakdown at high power near the ring at different potentials. Hence, we propose a new spurious mode suppressing design leveraging a thickened active ring in lithium niobate (LN), maintaining high $Q$ and $k^2$ while also reducing resistance at resonance ($R_r$), and mitigating breakdown effects.

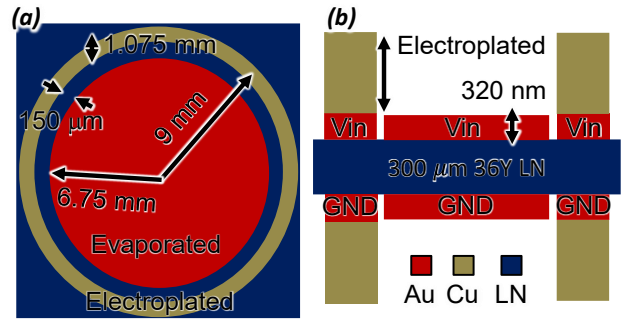

Fig. 1 (a) Top view of the device design. (b) Cross-sectional view of the device design.

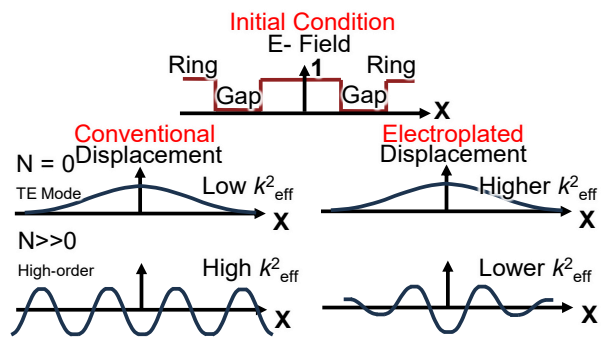

Fig. 2 Electrical excitation scheme, its effect on $k^2_{eff}$ and the effects of an electroplated ring on the normal mode displacement.

## I. INTRODUCTION

Acoustic resonators pose as compact alternatives for magnetic inductors to shift the paradigm of power conversion [1], [2], [3], [4], [5], [6]. While inductors are fundamental components in modern electronics, their inherent drawbacks – being bulky and lossy – pose significant challenges to the advancement of next-generation electronic systems [7]. These limitations are further compounded by the sub-linear performance scaling of inductors with frequency, creating a bottleneck in high-frequency technology. Hence, while other technologies mature, the inductor remains a major limitation for device efficiency and its footprint. This bottleneck has far-reaching implications, impacting the development of crucial technologies that shape daily life and contribute to global sustainability. The effort to shift from alternating current (AC) power conversion to direct current (DC), is a prime example [8]. This effort aims to dramatically bolster power conversion efficiency, leading to a range of far-reaching benefits [9]. However, the obtainable power conversion efficiency is largely limited by the inherent loss in the inductor utilized in the power circuit [10]. To overcome this issue, recent efforts have demonstrated that acoustic resonators pose as promising alternatives for lossy magnetic inductors [5], [11], [12], [13], [14], [15], [16], [17]. They leverage the direct piezoelectric effect to resonate with both a low impedance ($f_s$), and a high impedance resonance ($f_p$). Between $f_s$ and $f_p$, they are nearly purely inductive and can effectively be used as one. At the same time, acoustic resonators are incredibly compact compared to electromagnetic (EM) devices at the same frequency – an intrinsic quality stemming from the low acoustic wave velocity. Moreover, a variety of different piezoelectric materials, such as aluminum nitride (AlN), scandium aluminum nitride (ScAlN), and lithium niobate (LN), have all been validated in radio frequency (RF) applications to possess low acoustic and EM loss, e.g., quality factors ($Q$) in the thousands are demonstrated at GHz frequencies [18], [19], [20]. Hence, acoustic resonators are compact and low-loss energy storage mechanisms, making them ideal candidates for power conversion systems.

The acoustic resonator's restricted inductive region between its series and parallel resonances constitutes a major bottleneck for its performance as an inductor. Therein lies the importance of the piezoelectric material – the bandwidth between $f_s$ and $f_p$ is determined by the intrinsic properties of the material, namely, its electro-mechanical coupling coefficient ($k^2$). As LN possesses a very large $k^2$ while also possessing low loss, it is well-suited for piezoelectric power conversion [2], [21], [22], [5]. However, materials with a large available $k^2$ usually suffer from the emergence of spurious modes between $f_s$ and $f_p$, which further diminish the inductive region by introducing large spikes in device resistance [23], [24]. The suppression of such modes is a topic of ongoing research, and the key design criteria for a high $k^2$ platform for piezoelectric power conversion. Prior work has partially addressed this issue by the introduction of various design guidelines tailored to the material and the mode of interest, but can only provide a limited spurious-free region [25], [26]. We have previously demonstrated that a separated grounded ring on LN operating in the first order symmetric lamb mode (S1), maintains optimal device performance with spurious mode suppression [27]. Using this result, we synthesized a 3.2 kW DC-DC power converter with 97.9% peak efficiency [28]. However, we found that the grounded ring design suffers from a significant issue – it has been shown to experience voltage breakdown at high power near the gap at different potentials. Hence, we propose an alternative - a spurious mode suppressing design for LN resonators operating in the S1 mode while leveraging a thickened active ring, maintaining high $Q$ and $k^2$ while also reducing resistance at resonance ($R_r$), and most importantly, mitigating breakdown effects.

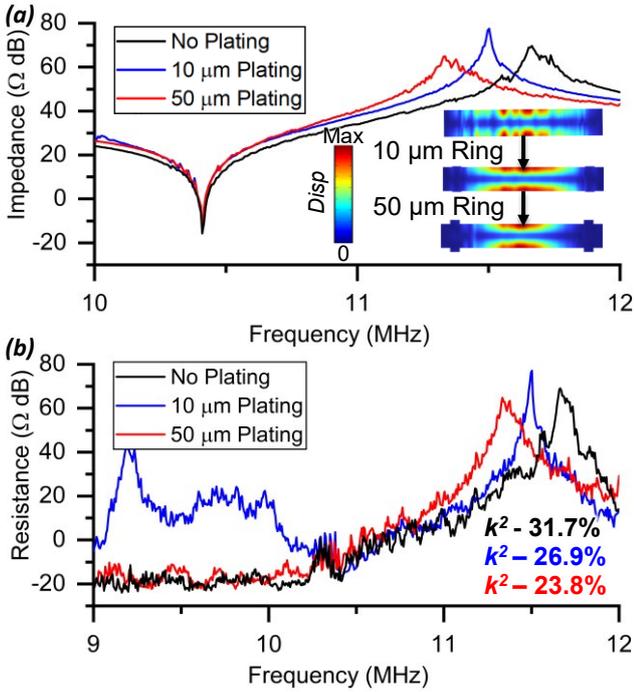

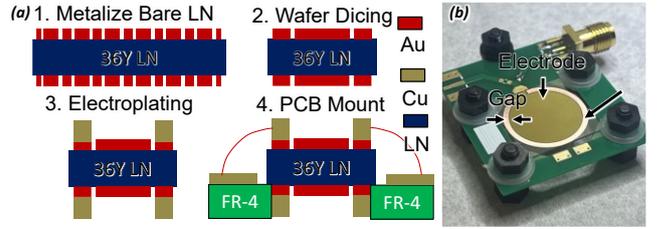

Fig. 3. (a) Simulated admittance indicating mode confinement and loss in $k^2$ as a function of electroplating thickness. (b) Simulated resistance, indicating spurious mode suppression.

## II. DESIGN AND SIMULATION

To maximize the inductive behavior of an acoustic resonator for power conversion, it should operate in a vibrational mode that possesses the highest electro-mechanical coupling ($k^2$) [2]. Though LN can obtain large $k^2$ from multiple modes, such as the thickness shear (TS), and thickness extensional (TE) modes, we will focus on TE mode LN resonators due to their unique dispersion [4].

To understand our current design procedure, we must briefly discuss our prior work on this topic. Previously, we have demonstrated TE mode, spurious–free acoustic resonators via a grounded ring [27]. Due to the strong piezoelectricity of LN and the layers of varying dispersion created by the active device, the gap, and the grounded ring, we were able to confine energy to the device's active area and suppress spurious modes between $f_s$ and $f_p$. However, it was found that this geometry suffered from breakdown effects as the grounded ring and the active area were at significantly different potentials when operating at high power. To combat this, we had to maintain the same potential within the entire device, eliminating the key component of our previous spurious mode suppression scheme – varying dispersion.

Fig. 4. (a) Fabrication flow-chart. (b) Images of the fabricated device.

Hence, in this work, our efforts were focused on maintaining spurious mode suppression with this fundamental change in the design.

Our new design can be seen in Fig. 1. Compared to conventional geometries, the device featured an active ring separated from center electrodes through a non-metalized gap. The spurious mode suppression mechanism here aimed to suppress all lateral overtones of the TE mode through mass-loading effects. The effect of the new electrical excitation structure on the resonant characteristics can be seen in Fig. 2. In this geometry, the exciting structure was no longer a straight line, rather, it was a train of step functions. Hence, the resonant tone was no longer the TE mode, but its lateral overtone. This occurred because the overtone possessed a higher effective $k^2$ ($k^2_{eff}$). In other words, the convolution between the exciting structure and the displacement profile of the lateral overtone yielded a higher value compared to its convolution with the fundamental tone. However, when the active ring was thickened via electroplated copper (Cu), the displacement magnitude in the ring was significantly damped due to mass loading, and the resonant mode progressed towards TE. This came at the cost of $k^2$ – the overall $k^2$ of the device moved closer to $k^2_{eff}$ of the fundamental TE tone, which was lower than the original $k^2$ by virtue of the design. This phenomenon was validated by simulations depicted in Fig. 3 (a)(b), where 10 μm Cu electroplating thickness was found to be optimal.

## III. FABRICATION

Device fabrication for this low-frequency prototype was performed via a bare 0.3 mm thick LN wafer with no substrate. 36Y LN cut was chosen due to its high available $k^2$ for the TE mode and low parasitic coupling compared to other LN crystal cuts. The wafer was first metalized on both top and bottom surfaces by 300 nm thick Au using electron beam evaporation, followed by wafer dicing. After the wafer was metalized and diced, the device was electroplated with 10 microns of Cu, followed by 40 microns of Cu to study the effects of mechanical damping on device performance. The fabricated devices were mounted on a PCB for measurement, where Au wire bonds were used to define the electrical connections. The fabrication flowchart and a device image can be seen in Fig. 4 (a)(b).

## IV. MEASUREMENT

Following device fabrication, devices were measured using a vector network analyzer (VNA). Attained data can be seen in Fig. 5 (a)(b), highlighting moderate $Q$, high $k^2$, and most importantly, spurious mode suppression via the active ring. This $Q$ is lower than our prior work on the same platform, hence we expect to gain back the $Q$ in consecutive experiments. The electrical response of the resonator as the electroplating goes on closely follows the simulation – $Q$ is

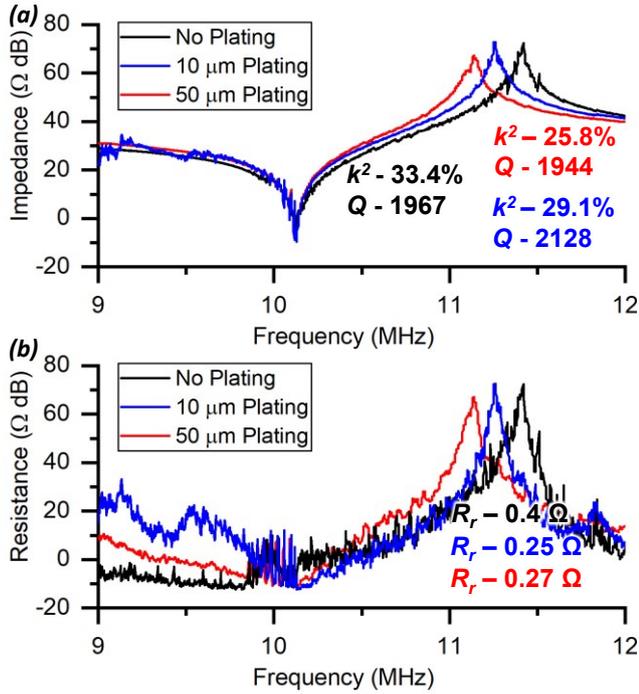

Fig. 5 (a) Measured impedance, depicting increase in $Q$ and spurious mode suppression at the optimal electroplating thickness. (b) Measured resistance, depicting spurious mode suppression and lowered resistance in-band at the optimal electro-plating thickness.

maintained and spurious modes are suppressed. Moreover, the resistance at resonance ($R_r$) is reduced, translating to lower device resistance between $f_s$ and $f_p$. All of this comes at the minor cost of $k^2$ – at the optimal electroplating thickness, the $k^2$ has degraded to 29.1% from the original 33.3%. This degradation is smaller than shown by simulations, which indicates that the electroplating recipe has overestimated the electroplated metal thickness.

At the optimal electroplating thickness, we see a slight increase in device $Q$. As we move past this optimal point, the device performance is affected negatively overall – resistance in-band is significantly increased, and the $k^2$ is further degraded. Interestingly, the device $Q$ is maintained, even at a 50 μm electroplating thickness, which indicates that the design effectively confines displacement to the device's active area. However, Cu still acts as a parasitic component, increasing resistance in-band. Hence, Cu electroplating leads to increasingly strong acoustic performance until the parasitic resistance of the electroplated metal becomes the dominant effect.

To numerically evaluate spurious-mode performance, we have adopted a metric commonly utilized in the state-of-the-art (SoA) – fractional spurious suppressed region [27]. This is a percentage that allows for a numerical evaluation of the effective bandwidth of a resonator for the purposes of power conversion. It is defined by the ratio between the frequency range in which the device resistance does not exceed twenty times the device resistance at resonance and the frequency range between $f_s$ and $f_p$. According to this evaluation, we have obtained a 61% spurious suppressed region. Compared with our previous achieved fractional spurious suppressed region of 62% via a grounded ring, we have maintained spurious mode suppression while avoiding potential breakdown issues when operating at high power. A more detailed comparison of our work to the SoA can be seen in table 1.

| Reference | Platform | $f_s$ (MHz) | FoM ($Q \cdot k^2$) | Spur. Supp. Region |
|---|---|---|---|---|
| Boles et al | PZT-Radial | 0.48 | 196 | 42.9% |
| Touhami et al | LN-TE | 6.28 | 9441 | N/A |
| Braun et al | LN-TE | 6.82 | 1212 | 3.38% |
| Nguyen et al | LN-TE | 5.94 | 1575 | 34.9% |
| Nguyen et al | LN-TE | 10.18 | 1200 | 62% |
| Stolt et al | LN-Radial | 0.3 | 9360 | N/A |
| **This Work** | **LN-TE** | **10.13** | **620** | **61%** |

Table 1. Comparison of our work to state-of-the-art, highlighting the maintenance of spurious mode suppression via an active ring.

## V. CONCLUSION

In this work, we have introduced a spurious mode suppression scheme that allows for the development of high $k^2$, high $Q$, spurious-free acoustic resonators for efficient DC-DC power conversion. Our approach addresses the limitation of previously demonstrated methods by maintaining a large fractional spurious suppressed region while preventing breakdown events at high power. More specifically, we present a near spurious-free acoustic device with a $k^2$ of 29.1% and a $Q$ of 2128 at 10.13 MHz. This result demonstrates the feasibility of more efficient, low-loss, and compact power conversion. These improvements can potentially redefine power transformers by overcoming a key technological bottleneck, paving the way for designs with high power density and a minimal footprint. Looking ahead, our platform offers exciting opportunities for the development of next-generation power converters, paving the way for energy efficiency and sustainability.

REFERENCES

[1]  J. D. Boles, J. J. Piel, N. Elaine, J. E. Bonavia, J. H. Lang, and D. J. Perreault, "Piezoelectric-Based Power Conversion: Recent Progress, Opportunities, and Challenges," Apr. 2022.
[2]  J. Boles, P. Acosta, Y. Ramadass, J. Lang, and D. Perreault, "Evaluating Piezoelectric Materials for Power Conversion," Nov. 2020.
[3]  W. Braun, Z. Tong, and J. Rivas-Davila, "Inductorless Soft Switching DC-DC Converter with an Optimized Piezoelectric Resonator," Mar. 2020.
[4]  K. Nguyen *et al.*, "Near Spurious-Free Thickness Shear Mode Lithium Niobate Resonator for Piezoelectric Power Conversion," *IEEE Trans. Ultrason. Ferroelectr. Freq. Control*, vol. 70, no. 11, Nov. 2023.
[5]  M. Touhami *et al.*, "Piezoelectric Materials for the DC-DC Converters Based on Piezoelectric Resonators," Nov. 2021.
[6]  J. D. Boles, "Power Electronics Meet Piezoelectrics: Converters, Components, and Miniaturization," Thesis, Massachusetts Institute of Technology, 2022.
[7]  W. G. Hurley and W. H. Wölfle, *Transformers and Inductors for Power Electronics: Theory, Design and Applications*. John Wiley & Sons, 2013.
[8]  M. Starke, L. M. Tolbert, and B. Ozpineci, "AC vs. DC distribution: A loss comparison," Apr. 2008.
[9]  M. A. Hannan, M. S. H. Lipu, P. J. Ker, R. A. Begum, V. G. Agelidis, and F. Blaabjerg, "Power electronics contribution to renewable energy conversion addressing emission reduction: Applications, issues, and recommendations," *Appl. Energy*, vol. 251, Oct. 2019,
[10] K. Górecki and K. Detka, "Influence of Power Losses in the Inductor Core on Characteristics of Selected DC–DC Converters," *Energies*, vol. 12, no. 10, May 2019,
[11] J. J. Piel, J. D. Boles, J. H. Lang, and D. J. Perreault, "Feedback Control for a Piezoelectric-Resonator-Based DC-DC Power Converter," Nov. 2021.

[12] E. Stolt *et al.*, "Fixed-Frequency Control of Piezoelectric Resonator DC-DC Converters for Spurious Mode Avoidance," *IEEE Open J. Power Electron.*, vol. 2, 2021,

[13] M. Touhami, G. Despesse, and F. Costa, "A New Topology of DC–DC Converter Based on Piezoelectric Resonator," *IEEE Trans. Power Electron.*, vol. 37, no. 6, Jun. 2022,

[14] B. Pollet, G. Despesse, and F. Costa, "A New Non-Isolated Low-Power Inductorless Piezoelectric DC–DC Converter," *IEEE Trans. Power Electron.*, vol. 34, no. 11, Nov. 2019,

[15] E. A. Stolt, W. D. Braun, and J. M. Rivas-Davila, "Forward-Zero Cycle Closed-Loop Control of Piezoelectric Resonator DC-DC Converters," Jun. 2022.

[16] B. M. Wanyeki, J. D. Boles, J. H. Lang, and D. J. Perreault, "Two-Stage Piezoelectric Resonator / Switched Capacitor DC-DC Converter," Oct. 2023.

[17] A. Saboor, Y. Hou, and K. K. Afridi, "Control Strategy for a Merged Switched-Capacitor Piezoelectric Resonator-Based DC-DC Converter Enabling Output Regulation at Fixed Frequency," Jun. 2024.

[18] J. Kramer *et al.*, "Thin-Film Lithium Niobate Acoustic Resonator with High Q of 237 and k2 of 5.1% at 50.74 GHz," May 2023.

[19] S. Gong, R. Lu, Y. Yang, L. Gao, and A. E. Hassanien, "Microwave Acoustic Devices: Recent Advances and Outlook," *IEEE J. Microw.*, vol. 1, no. 2, Apr. 2021,

[20] R. Lu and S. Gong, "RF acoustic microsystems based on suspended lithium niobate thin films: advances and outlook," *J. Micromechanics Microengineering*, vol. 31, no. 11, Sep. 2021,

[21] J. E. Bonavia, J. D. Boles, J. H. Lang, and D. J. Perreault, "Augmented Piezoelectric Resonators for Power Conversion," Nov. 2021.

[22] W.-C. B. Liu, G. Pillonnet, and P. P. Mercier, "An Integrated Dual-Side Series/Parallel Piezoelectric Resonator-Based DC–DC Converter," *IEEE J. Solid-State Circuits*, 2024,

[23] W. D. Braun *et al.*, "A Stacked Piezoelectric Converter Using a Segmented IDT Lithium Niobate Resonator," *IEEE Open J. Power Electron.*, vol. 5, 2024,

[24] E. Ng, J. D. Boles, J. H. Lang, and D. J. Perreault, "Piezoelectric Transformer Component Design for DC-DC Power Conversion," Jun. 2023.

[25] D. Rosén, J. Bjurström, and I. Katardjiev, "Suppression of spurious lateral modes in thickness-excited FBAR resonators," *IEEE Trans. Ultrason. Ferroelectr. Freq. Control*, vol. 52, no. 7, Jul. 2005.

[26] J.-H. Lee, C.-M. Yao, K.-Y. Tzeng, C.-W. Cheng, and Y.-C. Shih, "Optimization of frame-like film bulk acoustic resonators for suppression of spurious lateral modes using finite element method," Sep. 2004.

[27] K. Nguyen *et al.*, "Spurious-Free Lithium Niobate Bulk Acoustic Resonator for Piezoelectric Power Conversion," May 2023.

[28] E. Stolt, W. Braun, K. Nguyen, V. Chulukhadze, R. Lu, and J. Rivas-Davila, "A Spurious-Free Piezoelectric Resonator Based 3.2 kW DC–DC Converter for EV On-Board Chargers," *IEEE Trans. Power Electron.*, vol. 39, no. 2, Feb. 2024,